\documentclass{kluwer}
\usepackage[]{graphicx}

\begin{document}
\begin{article}
\begin{opening}

\title{Systematically Asymmetric Heliospheric Magnetic Field:
Evidence for a Quadrupole Mode and Non-axisymmetry with Polarity Flip-flops}

\author{K. \surname{Mursula}}
\institute{Department of Physical Sciences, University of Oulu,
Finland; email: Kalevi.Mursula@oulu.fi}

\author{T. \surname{Hiltula}}

\institute{Department of Physical Sciences, University of Oulu,
Finland}

\runningauthor{K. Mursula, and T. Hiltula}
\runningtitle{Hemispherical and Longitudinal Asymmetries}


\begin{abstract}

Recent studies of the heliospheric magnetic field (HMF) have detected
interesting, systematic hemispherical and longitudinal
asymmetries which have a profound significance for the understanding 
of solar magnetic fields.
The in situ HMF measurements since 1960s show that the heliospheric
current sheet (HCS) is systematically shifted (coned) southward during solar minimum
times, leading to the concept of a bashful ballerina.
While temporary shifts can be considerably larger, the 
average HCS shift (coning) angle is a few degrees, less than 
the $7.2^{\circ }$ tilt of the solar rotation axis.
Recent solar observations during the last two solar cycles verify these results 
and show that the magnetic areas in
the northern solar hemisphere are larger and their intensity weaker than 
in the south during long intervals in the late declining to minimum phase.
The multipole expansion reveals a strong quadrupole term which is oppositely
directed to the dipole term. 
These results imply that the Sun has a symmetric quadrupole S0 dynamo mode 
that oscillates in phase with the dominant dipole A0 mode.
Moreover, the heliospheric magnetic field
has a strong tendency to produce solar tilts that are 
roughly opposite in longitudinal phase. 
This implies is a systematic longitudinal asymmetry and leads to
a "flip-flop" type behaviour in the dominant HMF sector 
whose period is about 3.2 years. 
This agrees very well with the similar flip-flop
period found recently in sunspots, as well as with the observed ratio 
of three between the activity cycle period and the flip-flop period of sun-like stars.
Accordingly, these results require that the solar dynamo includes
three modes, A0, S0 and a non-axisymmetric mode.
Obviously, these results have a great impact on solar modelling.

\end{abstract}


\end{opening}

\section{Introduction}

Several studies during many decennia have examined possible longitudinal 
and hemispherical asymmetries in various forms of solar activity. 
E.g., there are well known prolonged periods when one of the solar hemispheres 
has dominated over the other in sunspot numbers (e.g., Carbonell, Oliver, and
Ballester, 1993; Oliver and Ballester, 1994), flare occurrence (e.g., Roy,
1977; Garcia, 1990)  or some other form of solar activity.  
However, the observed asymmetries in sunspots or other solar parameters
have not been found to be very conclusive, 
or to form any clear systematical pattern, e.g., 
in their relation to the 11-year solar activity cycle or 
to the 22-year solar magnetic cycle.  
Alas, the hemispheric and longitudinal asymmetries have not been
able to provide consistent input to solar theories
and, therefore, the significance of related studies has been quite marginal.

On the other hand, recent studies of similar longitudinal and
hemispherical asymmetries in the heliospheric magnetic field,
i.e, in the open solar magnetic field, have 
led to interesting results, revealing systematic and surprising 
properties of the global solar magnetic structure. 
First, observations during the first fast latitude scan in 1994-95
of the Ulysses probe found that the heliospheric current sheet
was shifted or coned southwards at this time 
(Simpson, Zhang, and Bame, 1996; Crooker {\it et al.},1997; Smith {\it et al.}, 2000).
More recently, using the 40-year series of in situ HMF observations,
it was shown (Mursula and Hiltula, 2003) 
that the southward shift or coning of the HCS is 
a common feature at least during the last four solar minima.
This feature has given the Sun a mnemonic nickname
of a "bashful ballerina" (Mursula and Hiltula, 2003) as
the solar ballerina is trying to 
push her excessively high flaring skirt (the HCS) 
downward whenever her activity fades away.

Second, it has recently been shown that the heliospheric magnetic field
has an interesting systematic behaviour in the occurrence of its
sectorial structure (Takalo and Mursula, 2002).
There is a strong tendency for the solar magnetic fields
to produce, in successive activations, solar tilts that have a
roughly opposite longitudinal phase from one
activation to another.
This implies a systematic longitudinal asymmetry in open solar
magnetic fields and leads to a "flip-flop" type behaviour 
in the dominant HMF sector.
The average period of one flip-flop during
the last 40 years is about 3.2 years (Takalo and Mursula, 2002),
in a good agreement with a more recent finding based on
a long series of sunspot observations (Berdyugina and Usoskin, 2003).

Here we review these recent developments in the structure and dynamics
of the heliospheric magnetic field
and discuss their implications to the solar theory and their
relation to similar, recent studies using solar surface observations.


\section{Hemispherical Asymmetry in HMF: Need for a Quadrupole S0 Mode}

In order to study the long-term hemispherical 
structure of the heliospheric
current sheet, Mursula and Hiltula (2003) used the hourly HMF data 
collected in the OMNI data set which covers in situ HMF observations at $1\ {\rm AU}$
since 1964. 
(The three HMF components Bx, By and Bz are given in the Geocentric Solar Ecliptic, GSE,
coordinate system in which the x-axis points from the Earth toward the Sun, z-axis
is perpendicular to the ecliptic plane and y-axis completes the
right-handed system, pointing roughly opposite to the Earth's
orbital velocity). 
For each hour, the HMF was divided into one of the two sectors, 
the toward or T sector consisting of field lines directed toward the Sun, or
the A sector directed away from the Sun.
Two different definitions were used to devide the HMF into two sectors: the plane
division and the quadrant division.
Because of the roughly $45^{\circ }$ winding angle of the HMF spiral at $1\ {\rm AU}$,
the T sector (i.e., the southern magnetic hemisphere) in the plane division
can simply be defined by the inequality Bx$\ >\ $By
(and the A sector by a reversed inequality).
Similarly, in the more restrictive quadrant division the T sector is defined by 
Bx$\ >\ $0 and By$\ <\ $0
and the A sector by Bx$\ <\ $0 and By$\ >\ $0.
Mursula and Hiltula (2003) calculated the total number of T and A sector hours for 
each 3-month season around the two high-latitude intervals 
(Spring = Feb--Apr; Fall$\ =$ Aug--Oct) and also for each full year,
as well as the corresponding normalized ratios 
T/(T+A) and A/(T+A)$\ =$ $1-{\rm T}/({\rm T}+{\rm A}$), i.e., 
the occurrence fractions of the two HMF sectors at any given time.

When, e.g., the fraction of the T sector in Fall
(when the Earth is at the highest northern heliographic latitudes) 
is plotted each year, a clear 22-year variation around the average of one half
is found so that the T sector dominates in Fall during the 
negative polarity minima (e.g., in the 1960s and 1980s), while
the A sector dominates in the positive minima (e.g., in the 1970s and 1990s).
This reflects the dominantly dipolar structure of the solar magnetic field 
around solar minima with dominant field polarity in either hemisphere
alternating from one cycle to another.
In the case of HMF this leads to the alternating dominance of one HMF sector
in Fall and Spring, the so called Rosenberg-Coleman (R-C) rule 
(Rosenberg and Coleman, 1969). 
Mursula and Hiltula (2003) quantified the R-C rule and found that the amplitude of the
22-year variation in the T/(T+A) fraction in Fall is $\pm 0.16$, implying
that the average ratio between the dominant and subdominant
sector occurrences in the northern heliographic hemisphere 
around solar minima is 1.94.
However, interestingly, the similar fraction in 
Spring, i.e., when the Earth is at the highest southern heliographic 
latitudes, was found to be significantly smaller, about $\pm 0.11$, implying
that in Spring the dominant sector only appears about 56\% more often
than the subdominant sector.

Thus, although the R-C rule is separately valid in both  
solar (heliographic) hemispheres, there is a systematic 
difference in the latitudinal HMF structure between the two hemispheres
so that the dominance of either HMF sector is 
systematically stronger in the northern than southern heliographic hemisphere.
This difference can be studied by the normalized ratio (T--A)/(T+A), i.e., the
difference in the fractional occurrence of T and A sectors.
The annual (or equinoctial) average of this ratio
can reveal the possible dominance of either magnetic hemisphere
during one year (or only at high heliographic latitudes) and, 
thereby, the possible north-south asymmetry of the HCS during that year.
Figure 1 depicts this ratio for a number of choices and shows that, 
despite some scatter (which is mostly not random but 
due to significant short-term variations, see later),
there is a systematic 22-year baseline oscillation in the dominant magnetic hemisphere.
Accordingly, the HMF sector prevalent in the northern heliographic hemisphere 
(the A sector during positive polarity minima 
and T sector during negative polarity minima) is dominating during all solar minima.
This implies that the heliosheet at $1\ {\rm AU}$ is, on an average, shifted 
or coned toward the southern heliographic hemisphere during these times.

%
\begin{figure}
\centerline{\includegraphics[width=0.85\textwidth]{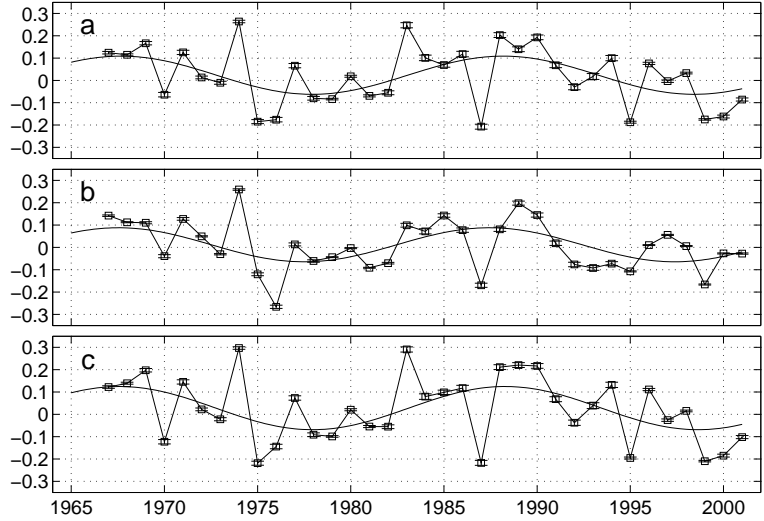}}
\caption{The (T-A)/(T+A) ratios in 1967-2001 together with the estimated errors and 
the best fitting sinusoids (Mursula and Hiltula, 2003).
a) Plane HMF division, Fall and Spring data only; 
b) Plane HMF division, all annual data;
c) Quadrant HMF division, Fall and Spring data only.}
\label{22_cycle}
\end{figure}
%
%
 
Note that the different choices of HMF sector definition (plane/ quadrant)
or data selection (full year/equinoxes) 
yield all a very similar 22-year oscillation in Figure 1. 
A typical amplitude of about 0.09 implies that, on an average, 
the HMF sector coming from the northern heliographic
hemisphere appears about 20\% more often around solar minima 
than the HMF sector from the southern hemisphere.
Since the R-C rule is, on an average, valid 
both in Fall and Spring, the average southward shift 
(coning) angle of the
heliospheric current sheet must be less than the 
$7.2^{\circ }$ tilt of the solar rotation axis. 
However, shifts can be temporarily much larger than this,
as also seen in Figure 1.
 
Further evidence for the southward shift of HCS
has recently been presented by Zhao, Hoeksema, and Scherrer (2004)
who have analysed the Wilcox Solar Observatory observations
of the solar magnetic field since 1976.
Using these observations and the (current-free potential field) 
source surface model, they calculated for each solar rotation
the total areas and average field strengths
of positive and negative polarity regions.
Figure 2 shows that the magnetic hemisphere dominant in the northern
solar hemisphere (negative polarity in mid-1980s, positive in mid-1990s)
has a systematically larger area than in the south for about three years
around the two solar minima included in the study.
There are also several shorter intervals of a few solar 
rotations where either of the two magnetic hemispheres is
temporarily dominating.  This shows that while a temporary north-south asymmetry in
magnetic hemispheres is indeed a quite typical situation,
a long-term asymmetry only appears in the late declining to minimum phase of the solar
cycle and is always depicting a larger area for that magnetic hemisphere
which is dominating the northern solar hemisphere.

%
\begin{figure}
\centerline{\includegraphics[width=\textwidth]{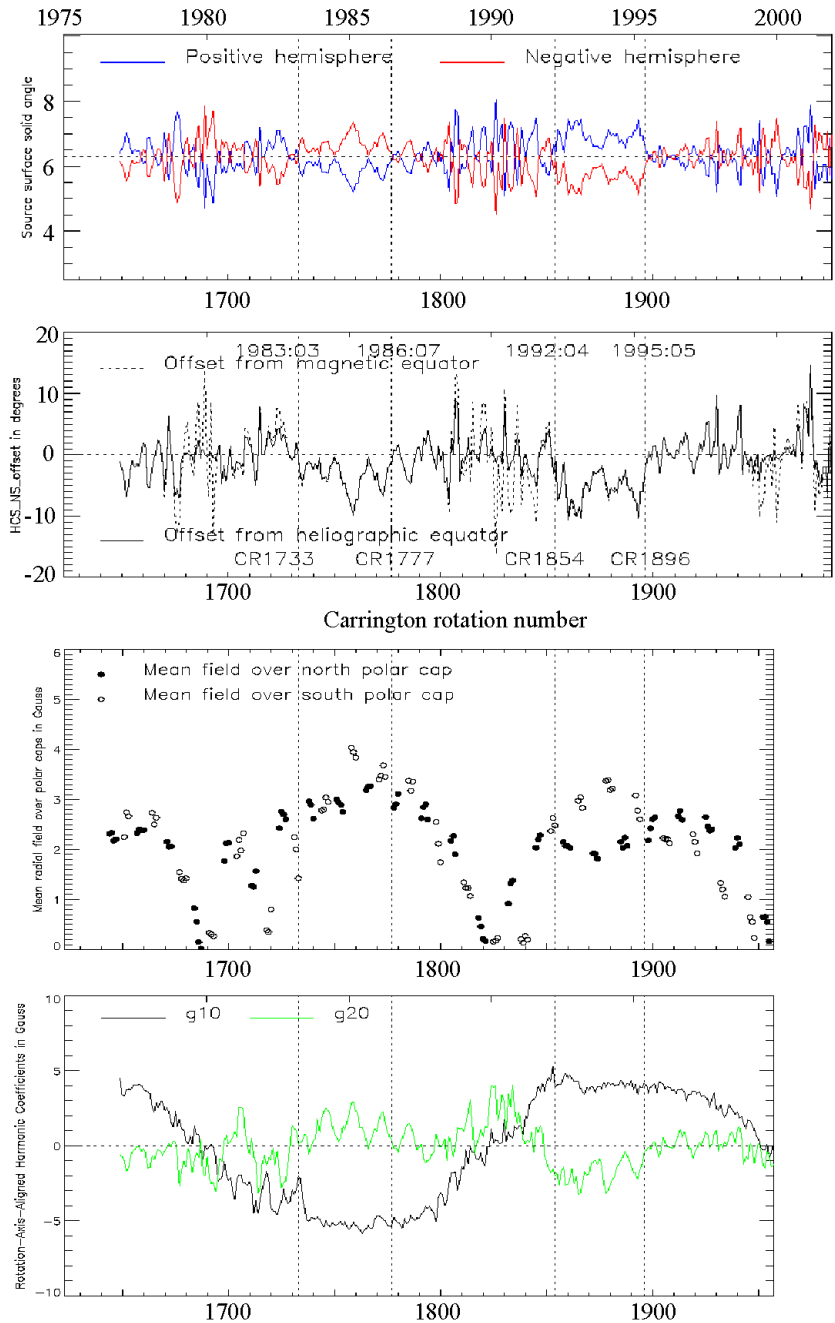}}
\caption{Solar magnetic field observations from WSO in 1976-2002 (Zhao, Hoeksema, and Scherrer, 2004).
a) Source surface areas of the two magnetic hemispheres (blue denotes positive
or away directed field, red negative);
b) Calculated HCS offset from heliographic (solid line) and magnetic
(dashed line) equator;
c) Mean radial field in the northern (full dot) and southern 
(open dot) polar cap;
d) Rotation axis aligned dipole (g10, black line) quadrupole 
(g20, green line) magnetic components.}
\label{Zhao}
\end{figure}
%
%
 
Figure 2 also shows that there is a
long-term southward shift in the calculated HCS location
during the  above mentioned intervals,
in a good overall agreement with the above general pattern
concluded from the HMF measurements (Mursula and Hiltula, 2003).
Even the magnitude estimated from the HMF observations
agrees very well with those depicted in Figure 2.
While the shift occasionally attains as large values
as $10^{\circ }$ or even more (note that this can also happen outside
the main asymmetry intervals),
the average shift during the three-year intervals
is about $3^{\circ }-5^{\circ }$, i.e., in a good agreement with the 
$7.2^{\circ }$ upper limit extracted from HMF observations.  

The larger area of magnetic field with that polarity which is dominating in the northern
solar hemisphere must, due to the equality of the total flux
of either polarity, be balanced by a larger intensity of
the magnetic field of opposite polarity which is dominating in the southern hemisphere.
Figure 2 shows that, indeed, the average intensity of 
the field in the southern polar cap is stronger than
in the northern polar cap roughly at the same 
times as the calculated long-term shift exists.
The asymmetry in the field strengths is slightly larger
in mid-1990s than in mid-1980s, in agreement with the slightly 
larger average HCS shift in mid-1990s (see Figure 2).
Hoeksema (1995) was among the first to note
that the average photospheric field strength in the northern polar cap 
is smaller than in the southern polar cap around solar minima.
Also, evidence has been found from the Ulysses magnetic field
measurements that the HMF intensity is stronger in the south
(Smith {\it et al.}, 2000).
Note also that the current-free potential field method used by Zhao et al. (2004)
implies that the asymmetry already exists in the photosphere and
that, e.g, no space currents are needed to explain the asymmetry.

A magnetic quadrupole term aligned with the solar rotation axis
has the same polarity in both polar regions contrary to 
the dipole term where the field at the two poles is oppositely oriented
(e.g., Bravo and Gonzalez-Esparza, 2000).
This difference is schematically depicted in Figure 3.
Thus, a significant quadrupole term can enhance the dipole term
at one pole and reduce it at the other pole, thus leading to 
a north-south difference in field strength, area and 
the related HCS asymmetry.
In order explain the observed higher field strength in the 
southern hemisphere (and the related larger area in the north and
the southward shifted HCS), the quadrupole term must be 
oriented opposite to the dipole term, as depicted in Figure 3.
Moreover, the quadrupole term must change its polarity in phase with 
the leading dipole term over the solar cycle since the 
north-south asymmetries and the HCS shift
remain oriented in the same direction from one cycle to another,
as observed with HMF measurements since 1960s (Mursula and Hiltula, 2003)
and with solar observations since 1980s (Zhao, Hoeksema, and Scherrer, 2004).
A symmetric quadrupole term is called the S0 mode
in solar dynamo theory. 
The present observations require that this mode must coexist
with the dominant dipole (A0) mode.

%
\begin{figure}
\centerline{\includegraphics[width=16pc]{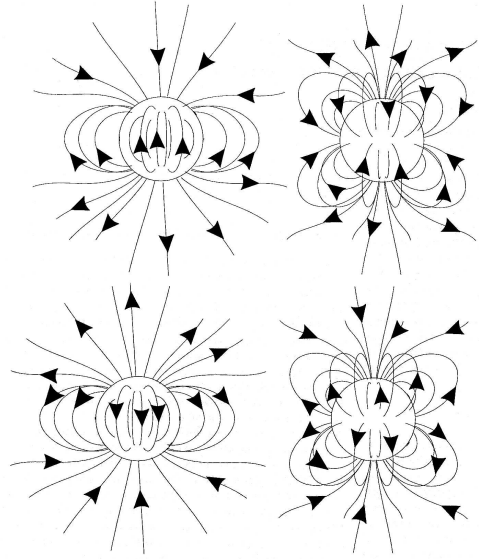}}
\caption{Schematic structure of symmetric dipole (left panels) and quadrupole 
(right panels) magnetic terms (Bravo and Gonzalez-Esparza, 2000). Polarities of the two
terms  in the same row are opposite.
Dipole polarity is negative (positive) in upper (lower) line.}
\label{dipole-quadrupole}
\end{figure}
%
%


\section{Longitudinal Asymmetry in HMF: Need for a Non-axisymmetric 
Mode with Rapid Flip-flops}

The solar wind and the heliospheric magnetic field have a strong tendency to repeat 
their current values after the solar rotation period of about 27 days.
This repetition reflects the existence of persistent, 
longitudinally asymmetric structures, such as, e.g., polar coronal holes
with equatorial extensions.
Persistent large scale magnetic fields also 
determine the inclination of the heliospheric current sheet
(the tilt of the solar magnetic field) and, thereby, the 
HMF sector structure observed, e.g., at $1\ {\rm AU}$.
It is known that the HMF sector structure typically prevails 
roughly the same for several solar rotations (e.g., Mursula and Zieger, 1996). 

In order to study the solar rotation related repetition
in HMF Takalo and Mursula (2002) 
calculated the autocorrelation function (ACF) of the
HMF Bx component up to lags of several tens of solar rotations.
Figure 4 shows that there is a strong tendency for HMF Bx to repeat 
its value with a decreasing probability (ACF amplitude) for about 9 solar rotations. 
This can be understood in terms of a slow decrease of the solar 
dipole tilt after some reconfiguration (tilt activation) 
produces an initial tilt value.
However, as first noted by Takalo and Mursula (2002),
after the node at about 10-11 rotations, the ACF amplitude 
(i.e., rotation periodicity) recovers again and reaches
an antinode at a lag of about 20-22 solar rotations. 
Moreover, after this first antinode the ACF amplitude decreases to the next  
node at a lag of about 35 rotations and increases again
to the next antinode at about 42-43 rotations, i.e., after some 3.2 years.
The long-term repetition of nodes and antinodes continues even thereafter,
as depicted in Figure 4c.

There are several important consequences of this node-antinode structure
of the ACF of the HMF in-ecliptic components.
(Very similar patterns are found both in Bx and By). 
First, this structure can not be produced if the subsequent 
tilt activations are random, but rather requires 
a considerable amount of phase coherence between such activations.
In particular, it implies that after one activation
has died out, the second activation develops at a high probability so that
its phase (longitude) is nearly opposite to the phase of
the first activation.
Moreover, the third activation has again a high probability 
to attain a tilt phase in the direction opposite to the 
second one, i.e., roughly reproducing the phase of the first activation.

%
\begin{figure}
\centerline{\includegraphics[width=0.9\textwidth]{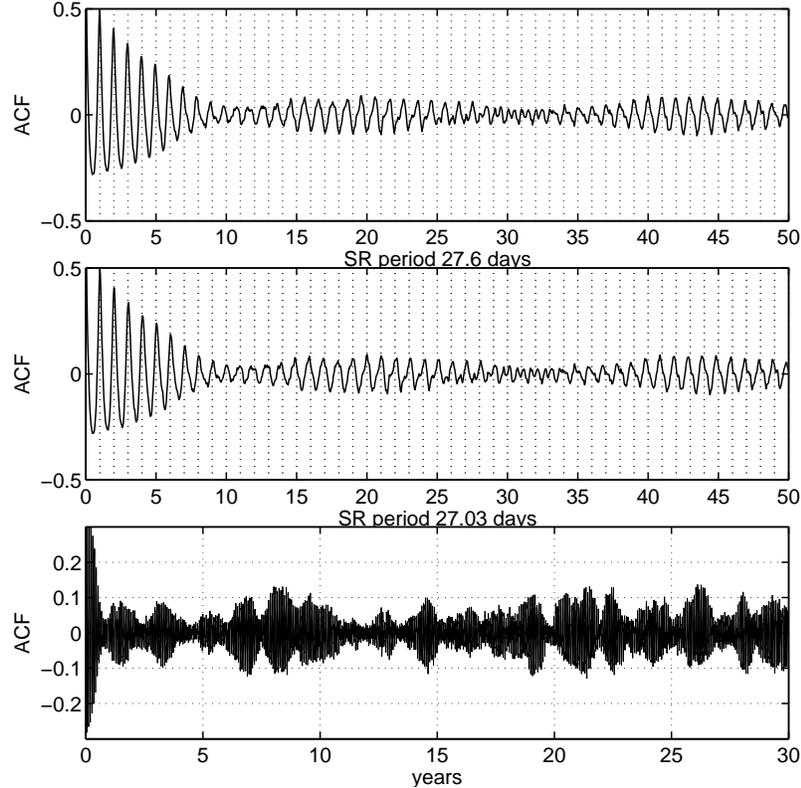}}
\caption{ACF of HMF Bx component for lags up to 50 solar rotations with 
a) multiples of the 27.6-day period marked with vertical dotted lines;
b) multiples of the  27.03-day period marked with vertical dotted lines.}
\label{ACF}
\end{figure}
%
%
 
Another important consequence of this node-antinode structure
relates to the average rotation period
of the large scale magnetic structures that produce the HMF.
In Figure 4a we have included multiples of the 27.6-day rotation period.
These multiples match very well to the first 10 ACF maxima  
and, after the first node, to the ACF minima (not maxima), 
reflecting the reversed phase in solar rotation periodicity (and solar tilt). 
The original phase is recovered again at the second antinode.  
This alternation of phase also continues with subsequent nodes.
Takalo and Mursula (2002) showed that such a node-antinode structure
of the ACF can be reproduced if the 27.6-day rotation periodicity is
phase/frequency modulated by the solar magnetic cycle.

Figure 4b depicts the same ACF together with 
multiples of a slightly shorter period of 27.03 days 
which was claimed (Neugebauer {\it et al.}, 2000) to be the most 
persistent rotation period in HMF sector structure. 
The 27.03-day multiples match with the ACF maxima during all antinodes. 
Thus the 27.03-day period retains the same phase throughout 
the  depicted time interval and
can not explain the observed nodal structure of the HMF in-ecliptic components. 
Accordingly, the slightly longer period of about 27.6 days
is, instead, the correct value for the most persistent rotation period 
of those solar magnetic structures responsible for the HMF.

The systematic preference of the solar tilt to repeat with 
opposite phase implies that there is a persistent, asymmetric pattern
in the large scale solar magnetic fields that cause such 
"active" longitudes in the tilt direction, 
in a quite similar way as has been found in the occurrence
of starspots (see, e.g., Berdyugina, 2004).
Moreover, the back and forth alternation of the tilt phase reminds of the 
alternation (so called "flip-flop") of the relative intensity
of the two active longitudes first found in starspots (Jetsu {\it et al.}, 1991) 
and later in sunspots (Berdyugina and Usoskin, 2003). 
As first found by Takalo and Mursula (2002), the period of
such a full flip-flop in HMF polarity (see Figure 4) takes
about 42-43 rotations, i.e., about 3.2 years.

It is interesting to note that the flip-flop period
in sunspots was recently estimated to be about 
3.6-3.8 years (Berdyugina and Usoskin, 2003), i.e., 
longer than the HMF flip-flop period. 
However, the sunspot analysis was based on a
much longer record which includes many weak and long cycles at the turn of the
20th and 21st centuries, while the HMF study only includes data from the recent
highly active and short cycles.
Thus, in fact, such a difference in the flip-flop period 
between the two studies is even expected if they
indeed reflect the same solar processes. 
This similarity is valid even more quantitatively.
Stellar observations suggest that the ratio between 
the stellar activity cycle and the flip-flop
cycle for sun-like stars is typically three (Berdyugina, 2004). 
Thus, multiplying the observed HMF flip-flop period by three gives about 10 years
which is indeed close to the average cycle length during the recent decennia.
Similarly, the longer sunspot flip-flop period multiplied by three gives 11 years
which is close to the long-term averaged sunspot cycle length. 
These results strongly suggest that the active longitudes
and their dynamics (flip-flops) have the same origin both 
for the large scale solar magnetic structures 
responsible for solar tilt and HMF, as well as for those producing sunspots.
Moreover, these results give further evidence for the above mentioned
stellar observation of a fixed ratio of three between the cycle
and flip-flop periods because they suggest that this 
ratio remains the same even if the cycle length changes 
significantly, as it has done in the Sun during the last 150 years.
These results also clearly demonstrate the need in the Sun for
a non-axisymmetric dynamo mode and thereby give another 
important constraint for solar dynamo modelling.

Finally, we would note that the development
of the HCS north-south asymmetry (the 22-year oscillation in Figure 1) is not 
sinusoidal but contains significant fluctuations around the trend.
There are large deviations from the sinusoidal pattern in the form of bipolar 
type fluctuations, e.g., in 1974-76.
Such bipolar fluctuations last typically 3-4 years and
are probably be related to the above discussed 
flip-flop periodicity in HMF sector structure. 
This and other connections between the hemispherical asymmetry 
(the bashful ballerina) and the longitudinal asymmetry 
will be studied in more detail in future publications.


\section{Conclusions}

The HMF observations since 1960s show that the 
heliospheric current sheet is systematically 
shifted or coned southward during solar minimum times (Mursula and Hiltula, 2003).
While temporary shifts are considerably larger, the 
average HCS shift (coning) angle was found to be smaller than 
the $7.2^{\circ }$ tilt of the solar rotation axis from the ecliptic.
These results have been verified by WSO observations of the solar
magnetic field for the last two solar cycles (Zhao, Hoeksema, and Scherrer, 2004).
While the areas of open magnetic field were often found to be  
asymmetric on short time scales of a few solar rotations,
prolonged asymmetric periods of about three years were found in 
the late declining to minimum phase of the solar cycle.
During both minima included in WSO study, the magnetic areas in
the northern solar hemisphere were larger and the intensities weaker at these times.
The calculated HCS was shifted (coned) southwards by an average angle of about $3^{\circ
}-5^{\circ }$. A multipole expansion of the field shows that there is a strong,
rotation axis aligned quadrupole which is opposite to the dipole field. 
Accordingly, there is a need in solar dynamo theory for a symmetric 
quadrupole (S0) mode which is oriented oppositely
to the main dipole (A0) field and changes its polarity with the same
phase as the dipole field. 
Such a quadrupole field can explain the large scale hemispheric differences
in the average magnetic field areas and intensities, as well as
the observed HCS shift.

We have also shown that the large scale magnetic fields 
producing the heliospheric magnetic field are systematically 
asymmetric in longitude.
They have a strong tendency to produce, in successive activations, 
magnetic tilts that are always roughly opposite in 
longitudinal phase to the previous tilt.
This leads to a "flip-flop" type behaviour for the dominance of one HMF sector.
The period of such a flip-flop is about 3.2 years during the
last 40 years (Takalo and Mursula, 2002).
We noted that this agrees very well with the similar flip-flop
period found in sunspots (Berdyugina and Usoskin, 2003), 
and supports the ratio of three between the activity
cycle period and flip-flop period found for sun-like stars.
Obviously, these results require the inclusion of a non-axisymmetric 
mode in realistic dynamo theories, in addition to
the symmetric A0 and S0 modes.
 
Finally, we would like to note that there are further complications
to this picture since the streamer belt (i.e., solar wind distribution)
observed at $1\ {\rm AU}$ depicts an inconsistent behaviour, 
being systematically shifted 
toward the northern {\it magnetic} rather than southern
heliographic hemisphere (Mursula, Hiltula, and Zieger, 2002).
So, during negative solar minima both the HCS
and the streamer belt are shifted toward the heliographic south
but during positive solar minima they are oppositely shifted.
Note also that the Ulysses observations in 1994-1995 have
shown that the HCS was shifted downward while the streamer belt was simultaneously
shifted northward (Crooker {\it et al.}, 1997), in agreement with this unexpected
general pattern which needs to be explained by subsequent studies.


\acknowledgements {Financial support by the Academy of Finland 
is greatfully acknowledged. 
We are also grateful to NSSDC for OMNI data.}


\end{article}

\end{document}